\documentclass[12pt,letter]{article}
\usepackage[section]{placeins}
\usepackage[table]{xcolor}
\usepackage{lipsum}
\usepackage{multirow}
\usepackage{setspace}
\usepackage{caption}
\usepackage{subfigure}
 \usepackage[square,numbers]{natbib}
\usepackage{anysize}
\marginsize{1.25in}{1.25in}{1in}{1in}
\usepackage{authblk}
\usepackage{graphicx}
\usepackage{color}
\usepackage{url}
\usepackage{latexsym}
\usepackage{amsmath}
\usepackage{amssymb}
\usepackage{amsfonts}
\usepackage{multirow}
\usepackage{bbm}
\usepackage{syntonly}
\usepackage{titlesec}
\usepackage{epstopdf}
\usepackage{graphics}
\usepackage{epsfig}
\usepackage{bm}
\usepackage{booktabs,multirow,tabularx}

\definecolor{brcolor}{rgb}{0.8,0.0,0.4}

\newcommand{\D}{\mbox{$\cal D$}}
\titleformat{\section}
    {\centering\normalfont\Large\bfseries}{\thesection.}{1em}{}

\newcommand{\mydir}{./Figs/}
\def\UnderWiggleTemp{\the\catcode`\@}
\catcode`\@=11
\ifx\UnderWiggle@Loaded\relax
  \catcode`\@=\UnderWiggleTemp
    \else \let\UnderWiggle@Loaded=\relax \fi
\newbox\U@BoxA
\newbox\U@BoxB
\newdimen\U@DimenA
\def\U@DoUnderWiggle{
  \offinterlineskip
  \vtop{
    \hbox{\vbox{\copy0}}
    \vskip 1.2pt  
    \vbox to 0.4pt{
      \hbox to\wd0{\hss\char'176\hss}
      \vskip0pt minus 1fil
    }
    \vskip 0.4pt  
  }
}
\def\UnderWiggle#1{{%
  \ifmmode
    \mathchoice
      {\setbox0=\hbox{$\displaystyle #1$}\U@DoUnderWiggle}
      {\setbox0=\hbox{$\textstyle #1$}\U@DoUnderWiggle}
      {\setbox0=\hbox{$\scriptstyle #1$}\U@DoUnderWiggle}
      {\setbox0=\hbox{$\scriptscriptstyle #1$}\U@DoUnderWiggle}
  \else
    \setbox0=\hbox{#1}\U@DoUnderWiggle
  \fi
}}
\catcode`\@=\UnderWiggleTemp

\newcommand{\be}{\begin{eqnarray}}         \newcommand{\uw}{\UnderWiggle}
\newcommand{\ee}{\end{eqnarray}}           \newcommand{\ba}{\begin{eqnarray*}}
\newcommand{\ea}{\end{eqnarray*}}          
                
                \newcommand{\eps}{\epsilon}

 \newcommand{\bl}{\begin{lemma}}
\newcommand{\el}{\end{lemma}}              \newcommand{\bd}{\begin{definition}}
                 \newcommand{\ed}{\end{definition}}
               
 \newcommand{\bc}{\begin{corollary}}
\newcommand{\ec}{\end{corollary}}          \newtheorem{prop}{Proposition}
\newcommand{\bp}{\begin{prop}}             \newcommand{\ep}{\end{prop}}

\def\boldfacefake #1{%
    \hbox{%
        \mathsurround=0pt
        \hbox to 0.4pt{$#1$\hss}%
        \hbox to 0.4pt{$#1$\hss}%
        \hbox {$#1$}%
    }%
}

\newcommand{\twofslides}[4]
{
\hbox to\hsize{\hss
    \vbox{\psfig{figure=\mydir/#1,width=#3,height=#4}}\qquad
    \vbox{\psfig{figure=\mydir/#2,width=#3,height=#4}}
    \hss}
\vskip 0.0truein
}

\newcommand{\twof}[4]
{
\hbox to\hsize{\hss
    \vbox{\psfig{figure=\mydir/#1,width=#3,height=#4}}   \qquad
    \vbox{\psfig{figure=\mydir/#2,width=#3,height=#4}}
    \hss}
}
\newcommand{\twofigspec}[2]
{
\hbox to\hsize{\hss
    \vbox{\psfig{figure=\mydir/#1,width=2.7in,height=3.6in}}\qquad
    \vbox{\psfig{figure=\mydir/#2,width=2.7in,height=3.6in}}
    \hss}
\vskip -0.1truein
\hbox to\hsize{\hss
    \vbox{ \begin{center}\mbox{\footnotesize \hspace{0.0in} {(a)}
                     \hspace{2.5in} {(b)}  }  \end{center} }
    \hss}
}
\newcommand{\twofig}[2]
{
\hbox to\hsize{\hss
    \vbox{\psfig{figure=\mydir/#1,width=2.7in,height=2.7in}}\qquad
    \vbox{\psfig{figure=\mydir/#2,width=2.7in,height=2.7in}}
    \hss}
\vskip -0.1truein
\hbox to\hsize{\hss
    \vbox{ \begin{center}\mbox{\footnotesize \hspace{0.2in} {(a)}
                     \hspace{2.7in} {(b)}  }  \end{center} }
    \hss}
}
\newcommand{\twofigland}[2]
{
\hbox to\hsize{\hss
    \vbox{\psfig{figure=\mydir/#1,width=2.8in,height=1.8in}}\qquad
    \vbox{\psfig{figure=\mydir/#2,width=2.8in,height=1.8in}}
    \hss}
\vskip -0.1truein
\hbox to\hsize{\hss
    \vbox{ \begin{center}\mbox{\footnotesize \hspace{0.1in} {(a)}
                     \hspace{2.8in} {(b)}  }  \end{center} }
    \hss}
}
\newcommand{\twofigl}[2]
{
\hbox to\hsize{\hss
    \vbox{\psfig{figure=\mydir/#1,width=1.9in,height=1.9in}}\qquad
    \vbox{\psfig{figure=\mydir/#2,width=1.9in,height=1.9in}}
    \hss}
\vskip -0.1truein
\hbox to\hsize{\hss
    \vbox{ \begin{center}\mbox{\footnotesize \hspace{0.1in} {(a)}
                     \hspace{1.9in} {(b)}  }  \end{center} }
    \hss}
}
\newcommand{\twofigsq}[2]
{
\hbox to\hsize{\hss
    \vbox{\psfig{figure=\mydir/#1,width=2.3in,height=2.3in}}\qquad
    \vbox{\psfig{figure=\mydir/#2,width=2.3in,height=2.3in}}
    \hss}
\vskip -0.1truein
\hbox to\hsize{\hss
    \vbox{ \begin{center}\mbox{\footnotesize \hspace{0.1in} {(a)}
                     \hspace{2.3in} {(b)}  }  \end{center} }
    \hss}
}
\newcommand{\twofigo}[2]{
\hbox to\hsize{\hss
    \vbox{\psfig{figure=\mydir/#1,width=2.2in,height=2.6in}}\qquad
    \vbox{\psfig{figure=\mydir/#2,width=2.2in,height=2.6in}}
    \hss}
\vskip -0.1truein
\hbox to\hsize{\hss
    \vbox{ \begin{center}\mbox{\footnotesize \hspace{0.1in} {(a)}
                     \hspace{2.2in} {(b)}  }  \end{center} }
    \hss}
}
\newcommand{\twofigos}[2]{
\hbox to\hsize{\hss
    \vbox{\psfig{figure=\mydir/#1,width=2.1in,height=2.6in}}\qquad
    \vbox{\psfig{figure=\mydir/#2,width=2.1in,height=2.6in}}
    \hss}
\vskip -0.4truein
\hbox to\hsize{\hss
    \vbox{ \begin{center}\mbox{\footnotesize \hspace{0.1in} {(a)}
                     \hspace{2.1in} {(b)}  }  \end{center} }
    \hss}
}
\newcommand{\twosmall}[2]{
\hbox to\hsize{\hss
    \vbox{\psfig{figure=\mydir/#1,width=1.3in}} \hspace{0.8in}
    \vbox{\psfig{figure=\mydir/#2,width=1.3in}}
    \hss}
\vskip -0.1truein
\hbox to\hsize{\hss
    \vbox{ \begin{center}\mbox{\footnotesize \hspace{0.1in} {(a)}
                     \hspace{1.9in} {(b)}  }  \end{center} }
    \hss}
}

\newcommand{\threef}[5]
{
\hbox to\hsize{\hss
  \vbox{\psfig{figure=\mydir/#1,width=#4,height=#5}}%
  \hss}
\hbox to\hsize{\hss
  \vbox{\psfig{figure=\mydir/#2,width=#4,height=#5}}%
  \hss}
\hbox to\hsize{\hss
  \vbox{\psfig{figure=\mydir/#3,width=#4,height=#5}}%
  \hss}
}

\newcommand{\fourf}[6] 
{
\hbox to\hsize{\hss
    \vbox{\psfig{figure=\mydir/#1,width=#5,height=#6}}\qquad
    \vbox{\psfig{figure=\mydir/#2,width=#5,height=#6}}
    \hss}
\vskip 0.1truein
\hbox to\hsize{\hss
    \vbox{\psfig{figure=\mydir/#3,width=#5,height=#6}}\qquad
    \vbox{\psfig{figure=\mydir/#4,width=#5,height=#6}}
    \hss}
\vskip -0.1truein
}
\newcommand{\fourfig}[6]
{
\hbox to\hsize{\hss
    \vbox{\psfig{figure=\mydir/#1,width=#5,height=#6}}\qquad
    \vbox{\psfig{figure=\mydir/#2,width=#5,height=#6}}
    \hss}
\vskip -0.1truein
\hbox to\hsize{\hss
    \vbox{ \begin{center}\mbox{\footnotesize \hspace{0.1in} {(a)}
                     \hspace{#5} {(b)}  }  \end{center} }
    \hss}
\vskip 0.1truein
\hbox to\hsize{\hss
    \vbox{\psfig{figure=\mydir/#3,width=#5,height=#6}}\qquad
    \vbox{\psfig{figure=\mydir/#4,width=#5,height=#6}}
    \hss}
\vskip -0.1truein
    \vbox{ \begin{center}\mbox{\footnotesize \hspace{0.1in} {(c)}
                     \hspace{#5} {(d)}  }  \end{center} }
\hbox to\hsize{\hss
    \hss}
\vskip -0.1truein
}
\newcommand{\twofv}[4]
{
\hbox to\hsize{\hss
    \vbox{\psfig{figure=\mydir/#1,width=#3,height=#4}}
    \hss}
\vskip -0.1truein
\hbox to\hsize{\hss
    \vbox{\begin{center} \footnotesize{(a)} \end{center}}
    \hss}
\vskip 0.1truein
\hbox to\hsize{\hss
    \vbox{\psfig{figure=\mydir/#2,width=#3,height=#4}}
    \hss}
\vskip -0.1truein
\hbox to\hsize{\hss
    \vbox{\begin{center} \footnotesize{(b)} \end{center}}
    \hss}
\vskip -0.1truein
}
\newcommand{\sixfig}[8]
{
\hbox to\hsize{\hss
    \vbox{\psfig{figure=\mydir/#1,width=#7,height=#8}}\qquad
    \vbox{\psfig{figure=\mydir/#2,width=#7,height=#8}}
    \hss}
\vskip -0.1truein
\hbox to\hsize{\hss
    \vbox{ \begin{center}\mbox{\footnotesize {(a)}
                     \hspace{#7} {(b)}  }  \end{center} }
    \hss}
\vskip 0.1truein
\hbox to\hsize{\hss
    \vbox{\psfig{figure=\mydir/#3,width=#7,height=#8}}\qquad
    \vbox{\psfig{figure=\mydir/#4,width=#7,height=#8}}
    \hss}
\vskip -0.1truein
\hbox to\hsize{\hss
    \vbox{ \begin{center}\mbox{\footnotesize  {(c)}
                     \hspace{#7} {(d)}  }  \end{center} }
    \hss}
\vskip 0.1truein
\hbox to\hsize{\hss
    \vbox{\psfig{figure=\mydir/#5,width=#7,height=#8}}\qquad
    \vbox{\psfig{figure=\mydir/#6,width=#7,height=#8}}
    \hss}
\vskip -0.1truein
\hbox to\hsize{\hss
    \vbox{ \begin{center}\mbox{\footnotesize  {(e)}
                     \hspace{#7} {(f)}  }  \end{center} }
    \hss}
}

\newcommand{\threespec}[6]{
\hbox to\hsize{\hss
    \vbox{\psfig{figure=\mydir/#1,width=#4,height=#5}} \hspace{0.1in}
    \vbox{\psfig{figure=\mydir/ieq.eps,width=#6,height=#5}} \hspace{0.1in}
    \vbox{\psfig{figure=\mydir/#2,width=#4,height=#5}} \hspace{0.1in}
    \vbox{\psfig{figure=\mydir/iplus.eps,width=#6,height=#5}} \hspace{0.1in}
    \vbox{\psfig{figure=\mydir/#3,width=#4,height=#5}}
    \hss}
}
\newcommand{\sixf}[8]{
\hbox to\hsize{\hss
    \vbox{\psfig{figure=\mydir/#1,width=#7,height=#8}} \hspace{0.1in}
     \hspace{0.1in}
    \vbox{\psfig{figure=\mydir/#2,width=#7,height=#8}} \hspace{0.1in}
    \hspace{0.1in}
    \vbox{\psfig{figure=\mydir/#3,width=#7,height=#8}}
    \hss}
\vskip 0.1truein
    \hbox to\hsize{\hss
    \vbox{\psfig{figure=\mydir/#4,width=#7,height=#8}} \hspace{0.1in}
   \hspace{0.1in}
    \vbox{\psfig{figure=\mydir/#5,width=#7,height=#8}} \hspace{0.1in}
  \hspace{0.1in}
    \vbox{\psfig{figure=\mydir/#6,width=#7,height=#8}}
    \hss}
}

\newcommand{\onee}[3]  
{
\includegraphics[width=#2,height=#3]{\mydir/#1}    
}

\newcommand{\twoo}[6]  
{
\mbox{\subfigure{\includegraphics[width=#3,height=#4]{\mydir/#1}}\quad \quad
      \subfigure{\includegraphics[width=#5,height=#6]{\mydir/#2}} }
}

\newcommand{\threee}[5]  
{
\mbox{\subfigure{\includegraphics[width=#4,height=#5]{\mydir/#1}}\quad
      \subfigure{\includegraphics[width=#4,height=#5]{\mydir/#2}}\quad
      \subfigure{\includegraphics[width=#4,height=#5]{\mydir/#3}} }}

\newcommand{\fourr}[6]  
{
\mbox{\subfigure{\includegraphics[width=#5,height=#6]{\mydir/#1}}\quad
      \subfigure{\includegraphics[width=#5,height=#6]{\mydir/#2}} }
\mbox{\subfigure{\includegraphics[width=#5,height=#6]{\mydir/#3}}\quad
      \subfigure{\includegraphics[width=#5,height=#6]{\mydir/#4}} }
}

   {\VerbatimEnvironment
    \begin{Sbox}\begin{minipage}{10cm}\begin{Verbatim}}%
   {\end{Verbatim}\end{minipage}\end{Sbox}
    \setlength{\fboxsep}{8pt}\fbox{\TheSbox}}

\begin{document}
\title{\textbf{Gamma-Minimax Wavelet Shrinkage with Three-Point Priors}}

\author{Dixon Vimalajeewa and Brani Vidakovic}
\affil{{\small \emph{Department of Statistics, Texas A\&M University, College Station, TX}}}

\date{}
\maketitle


\begin{abstract}
 In this paper we propose a method for wavelet denoising of signals contaminated with Gaussian noise
when prior information about the
$L^2$-energy of the signal  is available.
Assuming  the independence model, according to which the
wavelet coefficients are treated individually, we propose
 a simple, level dependent shrinkage rules that turn out to be
  $\Gamma$-minimax for a suitable class of priors.

The proposed methodology is particularly well suited in denoising
tasks when the signal-to-noise ratio is low, which is illustrated by simulations on
the battery of standard test functions. Comparison to
some standardly used wavelet shrinkage methods is provided.

\vspace*{0.1in}
\end{abstract}

\section{Introduction}
In this introductory section we review fundamentals of $\Gamma$-minimax estimation, wavelet shrinkage, and Bayesian approaches to wavelet shrinkage.
\subsection{$\Gamma$-minimax theory}
$\Gamma$-minimax paradigm, originally proposed by Robbins (1951),
deals with the problem of selecting decision rules in tasks of statistical inference.
The $\Gamma$-minimax approach falls
between the Bayes paradigm, which selects procedures that work
well ``on average aposteriori'', and the minimax  paradigm, which guards
against least favorable outcomes, however unlikely. This approach has evolved
from seminal papers in the fifties (Robbins, 1951; Good, 1952) and
early sixties, through an extensive research on foundations and
parametric families in the seventies,  to a branch of Bayesian
robustness theory, in the eighties and  nineties. In this latter
stage the Purdue Decision Theory group took a prominent role; a comprehensive discussion of the $\Gamma$-minimax can be
found in Berger (1984, 1985).

The $\Gamma$-minimax paradigm  incorporates the
prior information about the statistical model by a family of plausible
 priors, denoted by $\Gamma,$ rather than by a single prior.
Elicitation of ``prior families" is often encountered in
practice. Given the family of priors, the  decision maker selects an action that is optimal with respect to the least
favorable prior in the family.

Inference of this kind is often interpreted in terms of a game theory.
The decision maker (statistician) is Player II.
Player I, an intelligent opponent to  Player II,
selects a prior from the family $\Gamma$ that is least favorable to Player II.
Player II chooses an action that will minimize his loss, irrespective
of what what was selected by Player I. The action of  Player II, as a function of observed data,
is referred to as the $\Gamma$-minimax action.

Formally, if $\D$ is a set of all decision rules and $\Gamma$
is a family of prior distributions over the parameter space
$\Theta$, then a rule $\delta^* \in \D$ is $\Gamma$-minimax  if
\be
\label{eq:gmmx}
\inf_{\delta \in \D} \sup_{\pi \in \Gamma} r(\pi,\delta)=\sup_{\pi \in \Gamma}r(\pi,\delta^*) = r(\pi^*,\delta^*),
\ee
where $r(\pi,\delta)=E^{\theta}\left[E^{X|\theta}{\cal
L}(\theta,\delta)\right]=E^{\theta}R(\theta,\delta)$ is the Bayes
risk under the loss ${\cal L}(\theta,\delta)$. Here $R(\theta, \delta)=  E^{X|\theta}_{\theta}{\cal
L}(\theta,\delta)$ denotes the frequentist risk of rule $\delta$, $\pi^*$ is the least favorable prior, and ${\cal
L}(\theta,\delta)$ is the loss function, usually the squared error loss, $(\theta-\delta)^2.$
Note that when
$\Gamma$ is the set of all priors, the $\Gamma$-minimax rule
coincides with minimax rule; when $\Gamma$ contains a single prior, then
the $\Gamma$-minimax rule coincides with Bayes' rule with respect to that prior.  When the decision
problem, viewed as a statistical game, {\it has a value}, that is, when $\inf_{\delta \in \D} \sup_{\pi \in \Gamma} \equiv   \sup_{\pi \in \Gamma}\inf_{\delta \in \D}$, then the
$\Gamma$-minimax solution coincides with the Bayes rule with
respect to the least favorable prior. For the interplay between the $\Gamma$-minimax and Bayesian paradigms, see Berger (1985).
A review on $\Gamma$-minimax estimation can be found in Vidakovic (2000).

\subsection{Wavelet shrinkage} \label{WS}
We consider a $\Gamma$-minimax approach to
the classical nonparametric regression problem
\be y_i=f(t_i)+
\sigma \varepsilon_i, \quad i=1,\dots, n,\label{problem} \ee
 where
$t_i$, $i=1, \ldots,n$, is a deterministic equispaced design on
$[0,1]$, the random errors $\varepsilon_i$ are i.i.d. standard
normal random variables, and the noise level $\sigma^2$ may, or may
not, be known. The interest is to recover the function $f$ from
the observations $Y_i$. Additionally, we assume that the unknown
signal $f$ has a bounded $L^2$-energy, hence it assumes values
from a bounded interval. After applying a linear and orthogonal
wavelet transform, model in (\ref{problem}) becomes
\be
c_{J_0,k}&=&\theta_{J_0,k}+\sigma \eps_{J_0,k},~~k=0,\ldots, 2^{J_0}-1, \nonumber \\
 d_{j,k}&=&\theta_{j,k}+\sigma \eps_{j,k}, ~~
j=J_0,\ldots, J-1, ~k=0,\ldots, 2^j-1, \label{wavemod}
\ee
 where
$d_{j,k}$ ($c_{jk}$), $\theta_{j,k}$ and $\eps_{j,k}$ are the
wavelet (scaling) coefficients (at resolution $j$ and location
$k$) corresponding to $y$, $f$ and $\varepsilon$, respectively; $J_0$ and $J-1$ are the
coarsest and finest level of detail in the wavelet decomposition. If $\epsilon$'s are i.i.d. standard normal, an arbitrary wavelet coefficient
from (\ref{wavemod})
can be modeled as
\be [d|\theta] \sim {\cal N} (\theta, \sigma^2),
\label{model} \ee
where, due to (approximate) independence of the
coefficients, we omitted  the indices $j,k$. The prior
information on the energy bound of the signal energy implies that a wavelet
coefficient $\theta$ corresponding to the signal part in (\ref{problem}) assumes its values in a bounded
interval, say $\Theta=[-m(j), m(j)],$ which  depends on the level $j$.

Wavelet shrinkage rules have been
extensively studied in the literature, but mostly when no additional
information on the parameter space $\Theta$ is available.
For implementation of wavelet methods in non parametric
regression problems we refer to Antoniadis et al.
(2001), where  the methods are described and numerically
compared.

\subsection{Bayesian model in the wavelet domain}
 Bayesian shrinkage methods in  wavelet domains have
received considerable attention in recent years, a review can be found in Rem\'enyi and Vidakovic (2013).
 Depending on a prior, Bayes' rules are shrinkage rules. The shrinkage process is defined as follows: A shrinkage
rule is applied in the wavelet domain and the observed  wavelet
coefficients $d$ are replaced by with their shrunken versions
$\hat{\theta}=\delta(d)$. In the subsequent step, by the inverse wavelet transform, coefficients are transformed
back to the domain of original data, resulting in a data smoothing.
The shape of the particular rule $\delta( \cdot )$ influences the denoising performance.

Bayesian models on the wavelet coefficients have showed to be capable of
incorporating some prior information about the unknown signal, such as
smoothness, periodicity, sparseness, self-similarity and, for some
particular basis (Haar),  monotonicity.

 The shrinkage is usually achieved by eliciting a single prior
distribution $\pi$ on the space of parameters $\Theta,$ and then
choosing an estimator $\hat{\theta}=\delta (d)$ that minimizes
the Bayes risk with respect to the adopted prior.

It is well known that most of the noiseless signals encountered in
practical applications have (for each resolution level) empirical
distributions of wavelet coefficients  centered around zero and
peaked at zero. A realistic Bayesian model that takes into account
this prior knowledge should consider a prior distribution
for which the prior predictive distribution produces a reasonable agreement with observations. A
realistic prior distribution on the wavelet coefficient $\theta$
is given by
\be \pi(\theta)=\epsilon_0 \delta_0 + (1-\epsilon_0)
\xi(\theta), \label{single} \ee
where $\delta_0$ is a point mass at
zero, $\xi$ is a symmetric and unimodal distribution on the
parameter space $\Theta$ and $\eps_0$ is a fixed parameter in
$[0,1]$, usually level dependent, that regulates the amount of
shrinkage for values of $d$ close to 0. Priors for wavelet
coefficients as in (\ref{single}) have been indicated in the early
1990's by Jim Berger and Peter M\"uller (personal communication), considered
in  Vidakovic and Ruggeri (2000), and Remenyi and Vidakovic (2013), among others.

It is however clear that specifying a single prior  distribution
$\pi$ on the parameter space $\Theta$ can never be done exactly.
Indeed the prior knowledge of real phenomena always contains uncertainty and multitude of prior distributions can
match the prior belief, meaning that on the basis of the partial
knowledge about the signal, it is possible to elicit only a
family of plausible priors, $\Gamma$. In a robust Bayesian
point of view the  choice of a particular rule $\delta$ should not
be influenced by the choice of a particular prior, as long as it
is in agreement with our prior belief. Several approaches have
been considered for ensuring the robustness of a specific rule,
$\Gamma$-minimax being one  compromise.

In this paper we incorporate prior information on the
boundedness of the energy of the signal (the $L_2$-norm of the
regression function).  The prior information on the energy bound
often exists in real life problems, and it can be modelled by the assumption that the parameter
space $\Theta$ is bounded. Estimation of a bounded normal mean has
been considered in Bickel (1981), Casella and Strawderman (1981),
Donoho et al. (1990)
(in the minimax setup) and in Vidakovic and DasGupta (1996) (in
the $\Gamma$-minimax setup). It is however well known that
estimating a bounded normal mean represents a challenging problem. In
our context, if the structure of the prior (\ref{single}) can be
supported by the analysis of the empirical distribution of the
wavelet coefficients, the precise elicitation of the distribution
$\xi$ cannot be done without some kind of approximation. Of
course, when prior knowledge on the energy bound is available,
then any symmetric  distribution supported on the
bounded set, say $[-m, m]$, can be a possible candidate for
$\xi(\theta)$.

Let $\Gamma$ denote the family
\be \Gamma=\{\pi(\theta) =
\eps \delta_0+(1-\eps_0) \xi(\theta), \xi(\theta)\in\Gamma_{S[-m,m]}\} \label{family}, \ee
 where $\Gamma_{S[-m,m]}$ is the class of all  symmetric
distributions supported on $[-m,m]$, $\delta_0$ is point mass
at zero, and $\epsilon$ is a fixed constant between 0 and 1. We also require that distribution $\xi$ does not have atoms at 0.

We consider two models, both assume that wavelet coefficients follow normal distribution (which is a statement about the distribution of the noise),
$d \sim {\cal N}(\theta, \sigma^2).$ In the Model I, the variance of the noise is assumed known, while in the Model II the variance is not known and is given a prior distribution.

The rest of the paper is organized as follows. Section 2 contains mathematical
aspects and results concerning the $\Gamma$-minimax rules.  An exact risk analysis of
the rule is discussed in Section 3. Section 4 proposes a sensible
elicitation of hyper-parameters defining the model. Performance of
the shrinkage rule in the wavelet domain and application to a
 data set are given in Section 5. In Section 6 we
summarize the results and provide discussion on possible
extensions. Proofs are deferred to Appendix.

\section{Three-point Priors and $\Gamma$-minimax Rules}
In this section we discuss $\Gamma$-minimax shrinkage rules that are Bayes' with respect to three point priors in two scenarios, when the variance of the noise is known (Model I), and when it is not known (Model II).

\subsection{Model I}

Let a wavelet coefficient $d$ be modeled as in (\ref{model}), $d\sim {\cal N}(\theta, \sigma^2), ~\sigma^2 ~\mbox{known.}$ In practice, $\sigma^2$ is estimated from the wavelet coefficients, usually using a robust estimator of variance from the coefficients in the finest level of detail. Without loss of generality  the variance may be assumed to be equal  1. The following result gives a $\Gamma$-minimax shrinkage rule.

\noindent {\bf Theorem 1.}  Let
\ba
 [d|\theta, \sigma^2] \sim {\cal N}(\theta, 1),\\
\ea
and
\be
\pi(\theta) \in \Gamma=\{ \epsilon \delta_0 + (1-\epsilon) \xi(\theta)\},\label{gamma}
\ee
where $\epsilon$ is fixed in $[0,1],$ and $\xi(\theta)$ is any distribution on $[-m, m]$ without atoms at 0, that is, with no a point-mass-at-zero component.
Then for $0 < m\ leq m^*$ the least favorable prior is
\ba
[\theta] \sim \pi(\theta)=\epsilon \delta_0 + \frac{1-\epsilon}{2} (
\delta_{-m} + \delta_m ).
\ea

\begin{figure}
\centering
\twoo{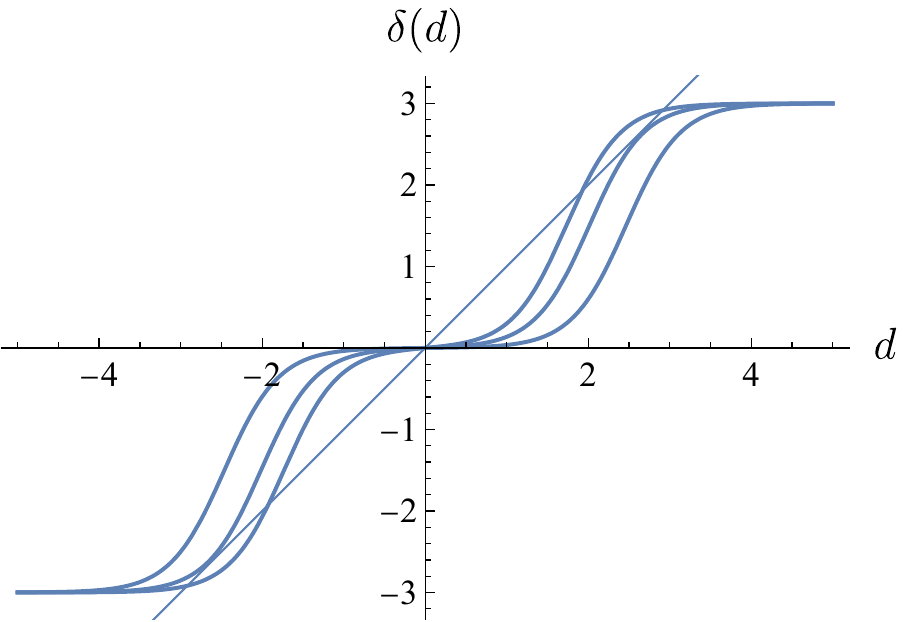}{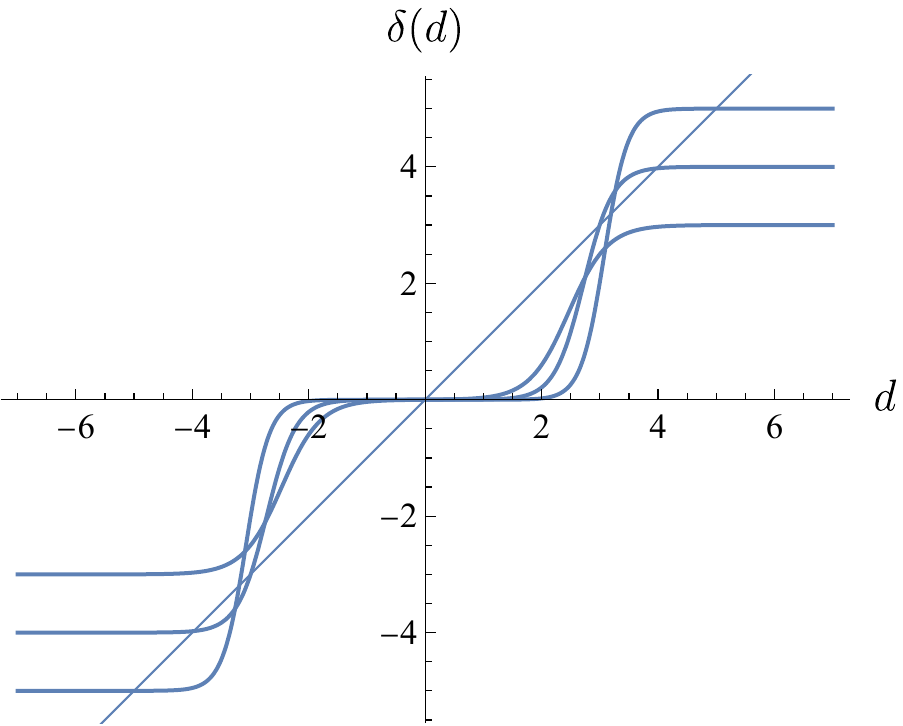}{2.5in}{2.0in}{2.5in}{2.0in}
\caption{$\Gamma$-Minimax Rule for Model I. Left: Rules for values for $m=3$ and $\epsilon=0.5, 0.7, $ and $0.9.$
Right: Rules for $\epsilon=0.9$ and $m=3, 4, $ and $5$.\label{fig:rules}}
\end{figure}

The  Bayes rule with respect to this prior,
\be
\label{eq:breI}
\delta_B(d) = \frac{ m \sinh(m d) }{\cosh(m d) + \frac{\epsilon}{1-\epsilon} e^{m^2/2}},
\ee
is the $\Gamma$-minimax rule. Figure \ref{fig:rules} shows the shinkage rule in (\ref{eq:breI}) for selected values
of parameters $\epsilon$ and $m$. Note that the rules heavily shrink small coefficients, but unlike traditional shrinkage rules,
remains bounded between $-m$ and $m.$ The values of $m^*$ are given in Table \ref{mstars}.

\begin{table}[h]
\centering
\begin{tabular}{|l|c|c|}
\hline
 $\epsilon$ & $m^*$~(Model I) & $m^*$~(Model II)\\
 \hline
 0.0 & 1.05674  &   0.81758  \\
 0.1 & 1.15020  &   0.91678  \\
 0.2 & 1.27739  &   1.05298  \\
 0.3 & 1.46988  &   1.25773  \\
 0.4 & 1.84922  &   1.52579  \\
 0.5 & 2.28384  &   1.74714  \\
 0.6 & 2.41918  &   1.91515  \\
 0.7 & 2.50918  &   2.05511  \\
 0.8 & 2.58807  &   2.19721   \\
 0.9 & 2.69942  &   2.40872   \\
 0.95 & 2.81605  &  2.63323   \\
 0.99 & 3.10039  &  3.24539   \\
 \hline
 \end{tabular}
 \caption{Values of $m^*$ for both models for different values of $\epsilon.$\label{mstars}}
 \end{table}

\subsection{Model II}
In Model II, the variance $\sigma^2$ is not  known and is given an exponential prior. It is well-known that the exponential distribution is an entropy maximizer
in the class of all distributions supported on $R^+$ with a fixed first moment.
This choice is noninformative, in a form of a maxent prior.

The model is:
\ba
&&[d|\theta, \sigma^2] \sim {\cal N}(\theta, \sigma^2),\\
&&[\sigma^2] \sim {\cal E}(\mu),~~~~~\left[ f(\sigma^2) = \frac{1}{\mu} \exp\left\{-\frac{\sigma^2}{ \mu}\right\}\right]
\ea
The marginal likelihood is double exponential as an exponential scale mixture of normals,
\ba
[d|\theta] \sim {\cal DE}\left(\theta, \sqrt{\frac{\mu}{2}}\right) ~~~~\left[ g(d|\theta, \mu) = \sqrt{\frac{1}{2 \mu}} \exp\left\{-\sqrt{\frac{2}{\mu}} |d - \theta|\right\}\right].
\ea

\begin{figure}
\centering
\twoo{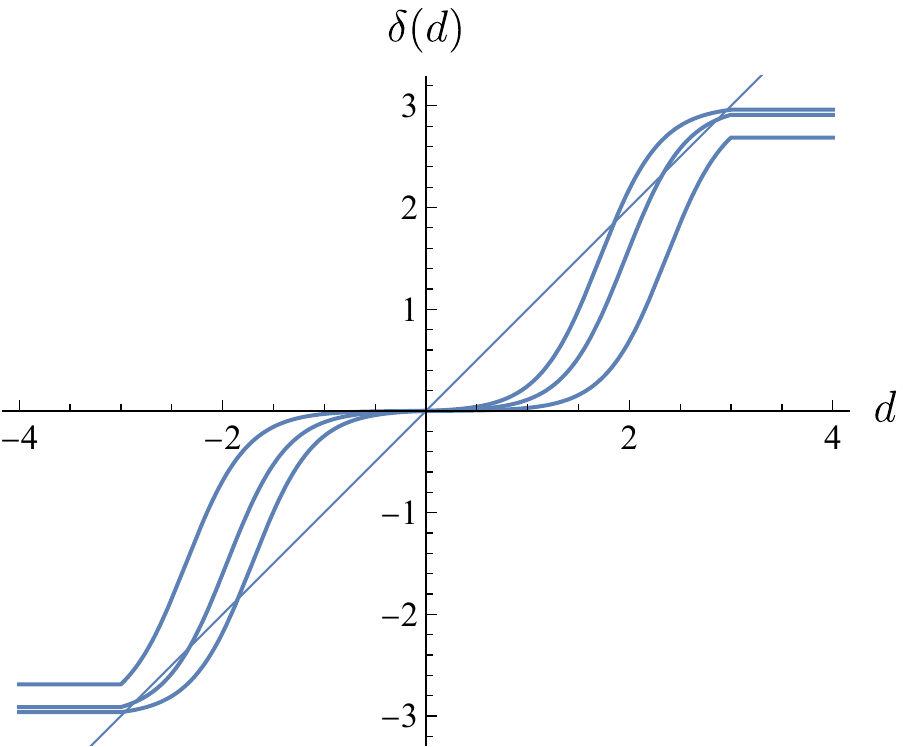}{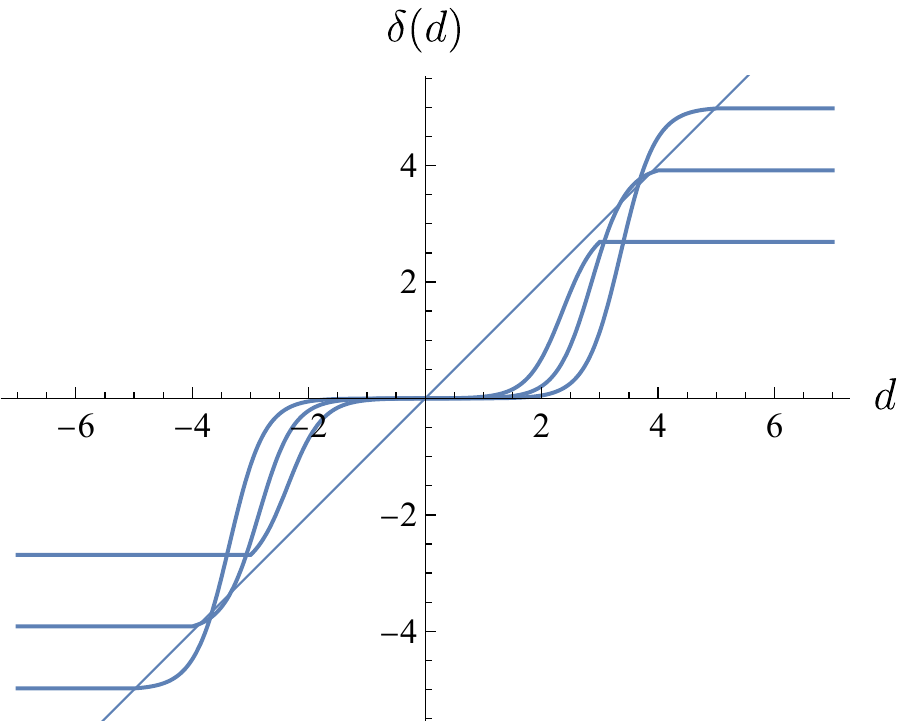}{2.5in}{2.0in}{2.5in}{2.0in}
\caption{$\Gamma$-Minimax Rule for Model II. Left: Rules for values or $\epsilon=0.5, 0.7, $ and $0.9.$
Right: Rules for $m=3, 4, $ and $5$.\label{fig:rulesde}}
\end{figure}

\noindent
{\bf Theorem 2.}
 If in Model II the family of priors on the location parameter is (\ref{gamma}), the
resulting $\Gamma$-minimax rule is:
\be
\label{eq:breII}
\delta_B(d) = \frac{ m \left( e^{-\sqrt{2/\mu} ~|d-m| }  - e^{-\sqrt{2/\mu} ~|d+m| } \right)}
{ e^{-\sqrt{2/\mu}~ |d-m| }  + \frac{2 \epsilon}{1-\epsilon}  e^{-\sqrt{2/\mu} ~|d| } + e^{-\sqrt{2/\mu} ~ |d+m| } },
\ee
is Bayes with respect to the least favorable prior
\ba
[\theta] \sim \pi(\theta)=\epsilon \delta_0 + \frac{1-\epsilon}{2} (\delta_{-m} + \delta_m ),
\ea
whenever $m \leq m^*.$  Figure \ref{fig:rulesde} shows the shinkage rule in (\ref{eq:breII}) for selected values of parameters $\epsilon$ and $m$.
The values of $m^*$ depend on $\epsilon$ and are given in Table \ref{mstars} for both models.
Sketches of proofs of Theorems 1 and 2 are deferred to Appendix.

\section{Risk, Bias, and Variance of $\Gamma$-Minimax Rules}

\begin{figure}
\centering
\onee{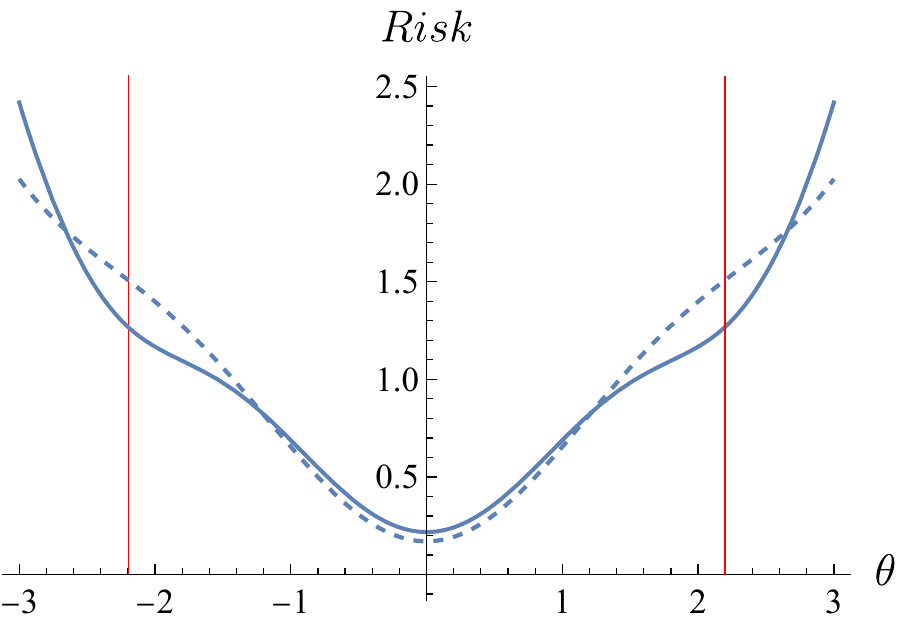}{2.5in}{1.7in}
\caption{Risk of $\Gamma$-Minimax rule for $\epsilon=0.8$ and $m=2.197.$ Dashed plots for Model I and solid for Model II.\label{FigRisks}}
\end{figure}

Frequentist risk of a rule $\delta$, as a function ot $\theta$, can be decomposed as a sum of two functions, variance and bias-squared,
\ba
R(\delta, \theta) = E^{d|\theta} (\delta(d)-\theta)^2 = E^{d|\theta}(\delta(d) - E^{d|\theta}(\delta(d) )^2 + (\theta -  E^{d|\theta}(\delta(d)) )^2.
\ea
To explore behavior of the two risk components in the context of Models I and II, we selected the risk of $\Gamma$-minimax rule for $\epsilon=0.8$ and $m=2.197.$
(Fig. \ref{FigRisks}).
This particular value of $m$ ensures that the rules are $\Gamma$-minimax (in fact $m=m^*$ for model II).
Note that $\delta$ in Model II shows smaller risk for values of $\theta$ in the neighborhood of $m$, while for $\theta$ close to 0, the risk of the rule from Model I is smaller.


\begin{figure}
\centering
\twoo{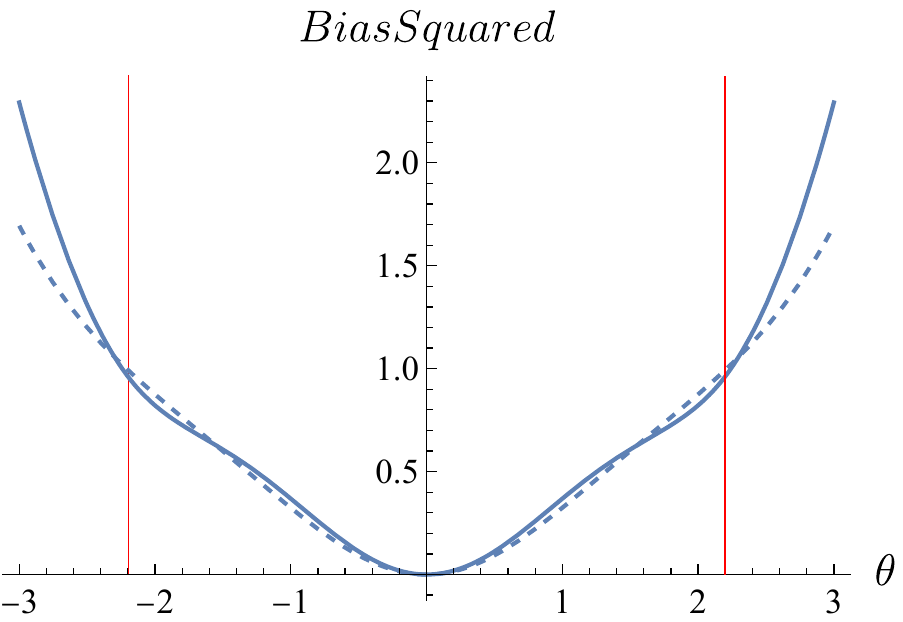}{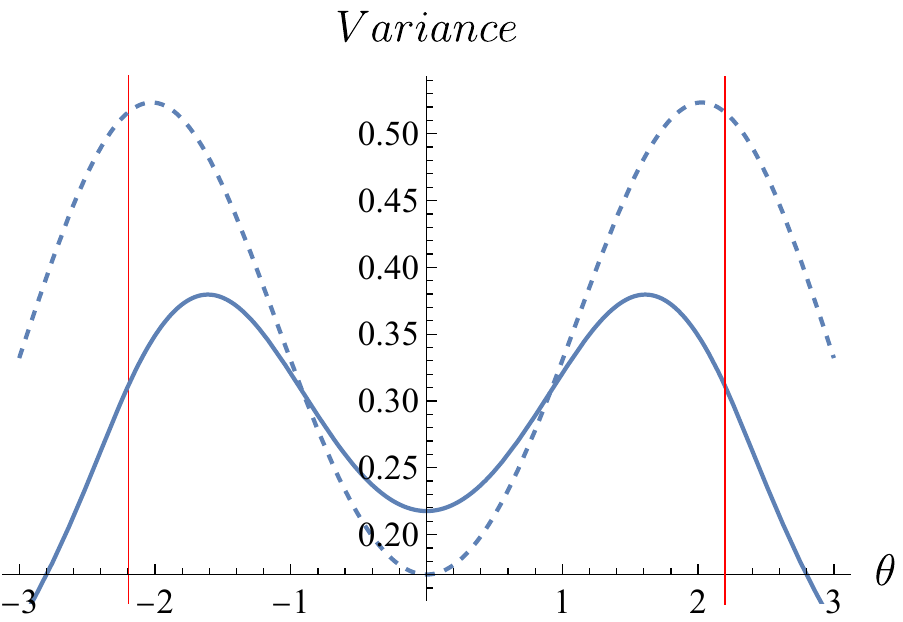}{2.5in}{1.7in}{2.5in}{1.7in}
\caption{ Risk components of $\Gamma$-minimax rule for $\epsilon=0.8$ and $m=2.197.$ Dashed plots for Model I and solid for Model II. Left: Bias-squared;
Right:  Variance    \label{BiasVar}}
\end{figure}

Similar, but less pronounced behavior is present in bias-squared function (Fig.\ref{BiasVar} Left Panel). Compared to Model I, the variance of $\delta$ in Model II is significantly smaller for values of $\theta$ in the neighborhood of $\pm m$, and larger for $\theta$ in the neighborhood of zero.
Preference in using either Model I or II depends on what size of signal part we are more interested. If there is more uncertainty about signal bound $m$, the rule form Model II is preferable. However, Model I has lower risk and both components of the risk in the neighborhood of 0. This translates to a possibly more precise shrinkage of small wavelet coefficients.

\section{Elicitation of Parameters}
The proposed Bayesian shrinkage procedures with three-point priors depend on three parameters, $m$, $\epsilon$ and $\mu$ that need to be specified. The criteria used for selecting these parameters are critical for effective signal denoising. We propose these hyper-parameters to be elicited in an empirical Bayes fashion, that is, dependent on the observed wavelet coefficients.
\begin{itemize}
    \item [(1)] {\bf Elicitation of $m$:} The bound $m$ in the domain of signal acquisition translates to level dependent bounds on the parameter $\theta$ in the wavelet domain. Given a data signal $\uw y=(y_1, \dots, y_n)$, the hyper-parameter  $m$ at the $j^{th}$ multiresolution level is estimated as
    \begin{equation}
    \label{eq:mj}
        m(j) = \hat{\sigma} \max(|y_i|) \left(\sqrt{2}\right)^{J-j},
    \end{equation}
    where $J = \log_2 n$ is the resolution level of the transformation, and $\hat \sigma$ an estimator of the noise size. The noise size is standardly estimated by a robust estimator of standard deviation that utilizes wavelet coefficients at the finest multiresolution level,
        \begin{equation}\label{sigma_hat1}
        \hat{\sigma} = \frac{ \text{median}(|d_{J-1,\bullet} - \text{median}(d_{J-1,\bullet})|)}{0.6745},
        \end{equation}
        where $d_{J-1,\bullet}$ represents all detail coefficients in the level $J-1.$   The multiple $\left(\sqrt{2}\right)^{J-j}$ in
         (\ref{eq:mj}) reflects the  increase
        of the extrema of absolute value of wavelet coefficients corresponding to a signal part in $j$ steps of the transform.

    \item [(2)] {\bf Elicitation of $\epsilon$:} This parameter controls the amount of shrinkage in the neighborhood of zero and overall shape of shrinkage rule. For the levels of fine details this parameter should be close to 1. Based on the  proposal for the $\epsilon$ given by Angelini and Vidakovic  (2004) and Sousa et al. (2021), we suggest a level-dependent   $\epsilon_0$ as follows:
    \begin{equation}\label{ep_0}
        \epsilon_0(j) = 1 - \frac{1}{(j - J_0 + l)^k},
    \end{equation}
    where $ J_0 \leq j \leq J-1$ and $k$ and $l$ are positive constants.

    As we indicated, $\epsilon$ should be close to one at the multiresolution levels of fine details, and then  be decreasing gradually for levels approaching the coarsest level (Angelini and Vidakovic, 2004). When $k$ and $l$ are large, $\epsilon$ remains close to one over all levels. This results in an almost noise-free reconstruction, but could result in over-smoothing. On the other hand, $l >1$ guarantees a certain level of shrinkage even at the coarsest level. Thus, the hyperparameters $l$ and $k$ should be selected with a care in order to achieve good performance for a wide range of signals. Numerical simulations guided us to suggest values $l \geq 6$ and $k = 2$ as reasonable choices. However, it is important to note that these parameters should depend on the smoothness of data signals and their size. We further discuss the selection of $l$ and $k$ for specific signals in Section \ref{simu_study}.

    \item [(3)] {\bf Elicitation of $\mu$:} This parameter is needed only for Model II. Since the prior on the noise level $\sigma^2$ is exponential and the prior mean is $\mu$, by moment matching we select $\mu$ as  $\hat \sigma^2$. A possible choice for  $\hat \sigma^2$ is a robust estimator as in (\ref{sigma_hat1}).

\end{itemize}

\section{Simulation Study}\label{simu_study}
\begin{figure*}[!t]
\centering
    \includegraphics[width= 1\linewidth]{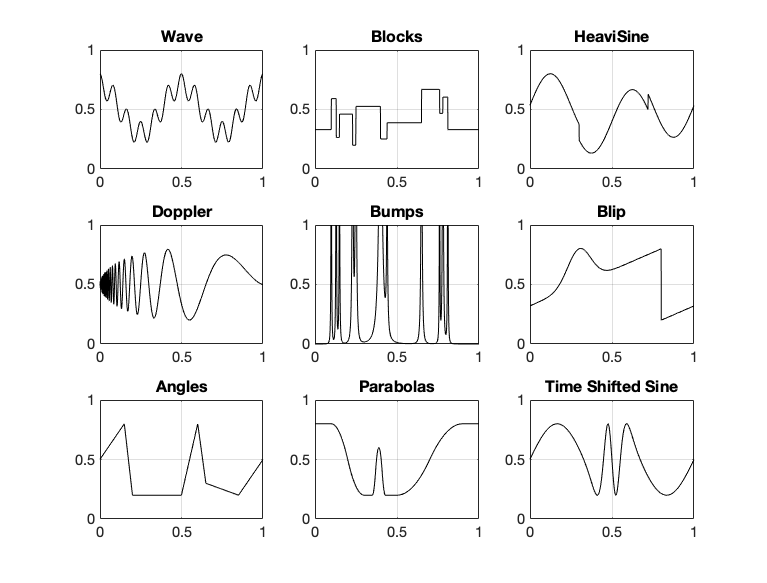}
    \caption{The nine test signals used in the simulation study.}
    \label{fig-1}
\end{figure*}

In the simulation study, we assessed the performance of the proposed shrinkage procedures on the battery of standard test signals. We used nine different test signals ({\tt step, wave, blip, blocks, bumps, heavisine, doppler, angles,} and {\tt parabolas}), which are constructed to mimic a variety of signals encountered in applications (Fig. \ref{fig-1}). As standardly done in literature, Haar and Daubechies six-tap  (Daubechies 6) were used for  Blocks and Bumps and Symmlet 8-tap filter was used for the remaining test signals. The shrinkage procedures are compared using the average mean square error (AMSE), as in (\ref{amse}).
All simulations were performed using   MATLAB software and toolbox \emph{GaussianWaveDen} (Antoniadis et al., 2001) that can be found at
\url{http://www.mas.ucy.ac.cy/~fanis/links/software.html}.

\begin{figure*}[!t]
\centering
 \includegraphics[width= 1\linewidth]{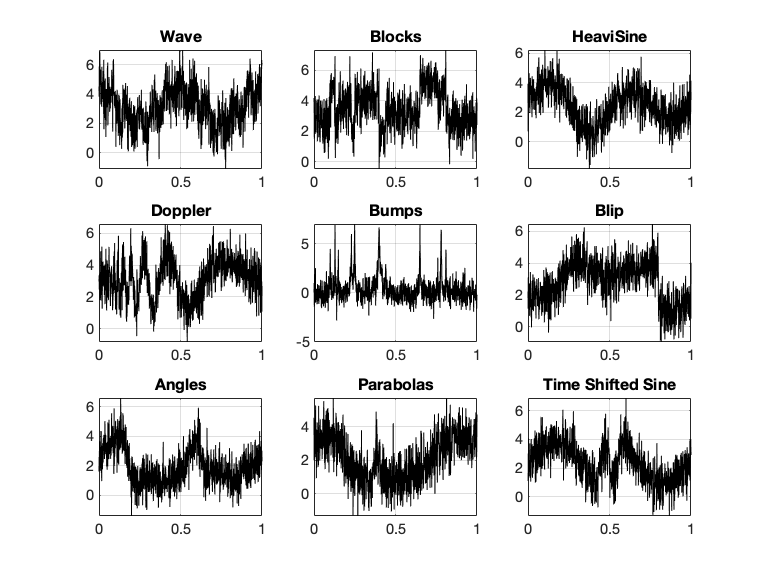}
        \caption{Noisy versions of the nine signals from Fig. \ref{fig-1}.}
\label{fig-2}
\end{figure*}

We generated noisy data samples of the nine test signals by adding normal noise with zero mean and variance $\sigma^2 = 1$. The signals were rescaled so that
$\sigma^2 =1$ leads to a prescribed SNR.
Each sample consisted of $n = 1024$ data points equally spaced on the interval $[0,1]$.  Figure \ref{fig-2} shows a noisy version of the nine test signals with SNR=1/4. Each noisy signal was transformed into the wavelet domain. After the  shrinkage   was applied to the transformed signal, the inverse wavelet transform was performed on the processed coefficients to produce a smoothed version of a signal in the original domain. The AMSE
was computed as
\begin{equation} \label{amse}
    AMSE(f(t)) = \frac{1}{n N}\sum_{j=1}^N \sum_{i=1}^n \left(f(t_i) - \hat{f}_j(t_i)\right)^2,
\end{equation}
where $f$ denotes the original test signal and $\hat{f_j}$ its estimator in the $j$-th iteration.
To calculate the average mean square error this process was repeated  $N = 100$ times.

\begin{figure*}[!t]
\centering
 \includegraphics[width= 1\linewidth]{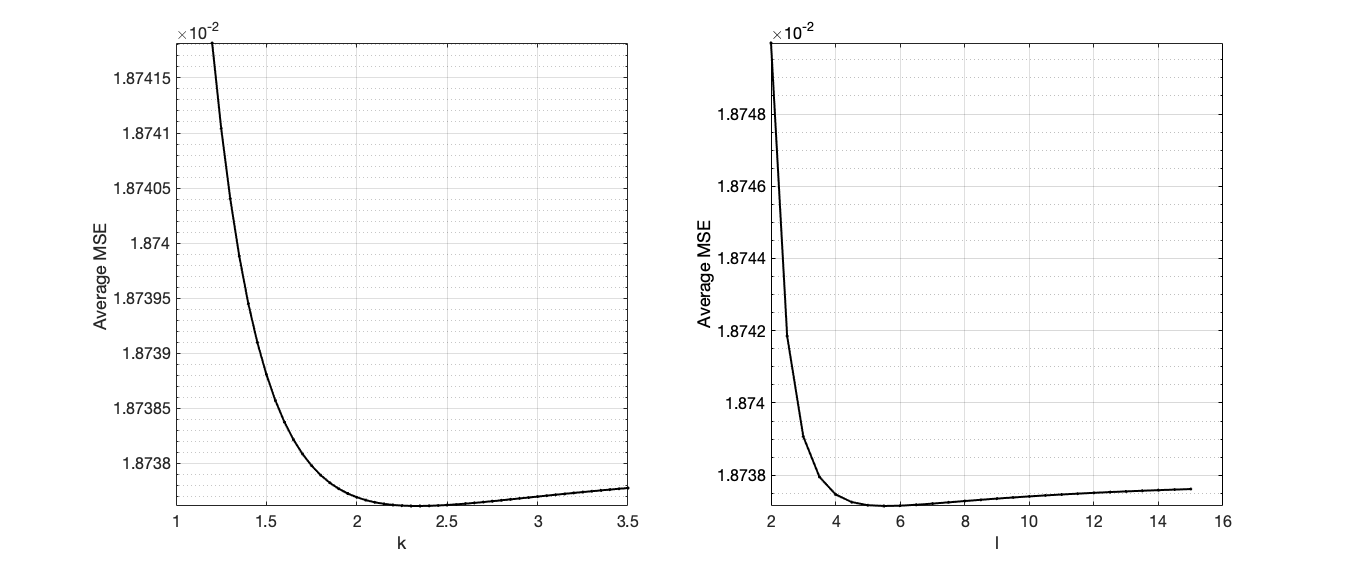}
        \caption{Change in average MSE as a function of hyper-parameters (a) $l$ and (b) $k$  exemplified on the Heavisine test signal, $SNR = 1/5$, and $n = 1024$.}
\label{fig-3}
\end{figure*}

\begin{figure*}[!t]
\centering
 \includegraphics[width= 1\linewidth]{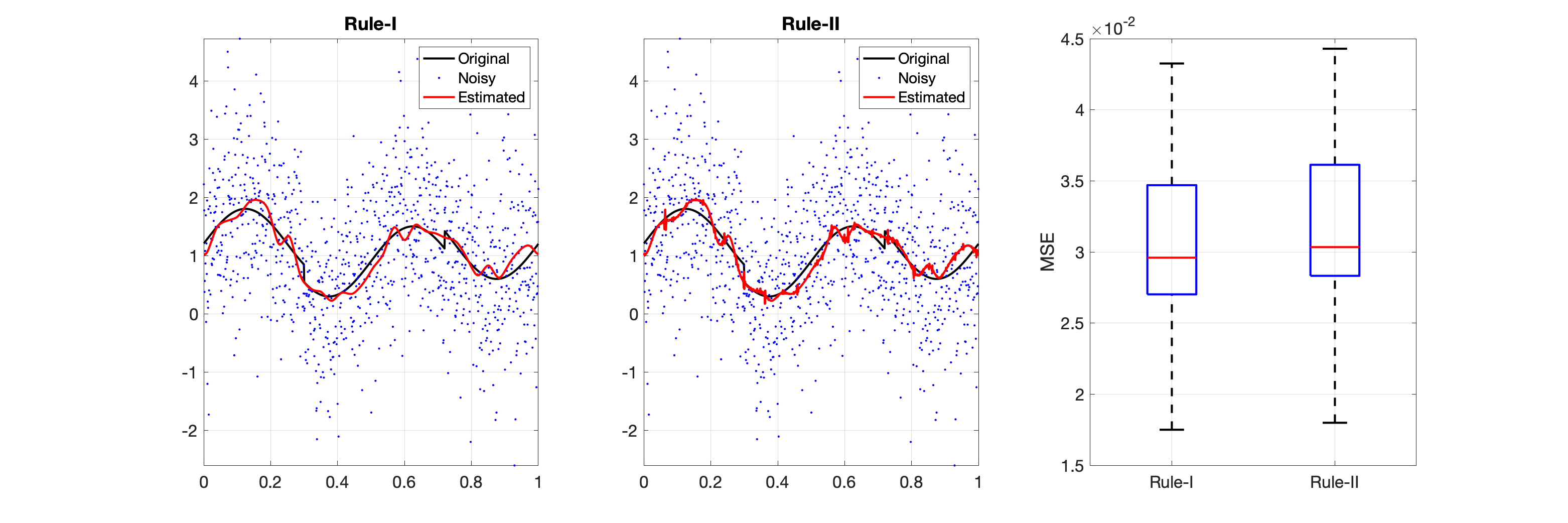}
        \caption{Estimation of {\tt heavisine} test signal: Estimations obtained by three-point priors with $ n = 1024, k = 2.4, l = 5.8$ and $SNR = 1/5$.}
\label{fig-4}
\end{figure*}

 The shrinkage procedure was applied to each test signal and the MSE was computed for a range of parameter values of $l$ and $k$. For example, Fig. \ref{fig-4} shows the average MSE obtained on the {\tt heavisine} test signal when SNR=1/5, and $l$ and $k$ vary in the range $l \in [2, 15]$ and $k \in [1.0, 3.5].$   . As evident from Fig. \ref{fig-4}, the estimator achieves its best performance for values $k \approx 2.4$ and $l \approx 5.8$. With these selected values of $l$ and $k$, Fig. \ref{fig-3} shows that the estimator is sufficiently close to the original test signal, even though the SNR is quite small. 

Based on our simulations, the optimal hyper-parameter values of $l$ and $k$  varied depending on the nature (e.g., smoothness) of test signal. For larger values of  $k$ and $l$, the estimator performs better for smooth signals. This is because the corresponding wavelet coefficients rapidly decay with the increase in resolution. However, larger values of $l$ and $k$ may not detect localized features in signals (e.g., cusps, discontinuities, sharp peaks), resulting in over-smoothing. For low values of SNR, the three-point priors estimator is more sensitive to hyper-parameter values of $l$ and $k$. When SNR increases, the estimation method shows better performance for most of the test signals with relatively small values of $l$ and $k$. Moreover, higher values of parameter $l$ and $k$ are preferred when the sample size is large.

In general, we  suggest that $k = 2.5$ and $l \geq 6$ as the most universal choice. The suggested values could, however, be adjusted depending on available information about the nature of signals.

\subsection{ Performance Comparison with Some Existing Methods}
We compared the performance of the proposed three-point prior estimator with eight existing estimation techniques. The selected existing estimation techniques include: Bayesian adaptive multiresolution shrinker (BAMS) (Vidakovic snd Ruggeri, 2001),   Decompsh (Huang and Cressie, 2000), block-median and  block-mean (Abramovich et al., 2002), hybrid version of the block-median procedure (Abramovich et al., 2002), blockJS (Cai, 1999), visu-shrink (Donoho and Johnstone, 1994), and generalized cross validation (Amato and Vuza, 1997). The first five techniques are relying on a Bayesian procedure and they are based on level-dependent shrinkage. The blockJS method uses a level-dependent thresholding,  while the visu-shrink and generalized cross validation techniques use a global thresholding method. Readers can find more details about these techniques in Antoniadis et al. (2001).

\begin{figure*}[!t]
\centering
 \includegraphics[width= 1\linewidth]{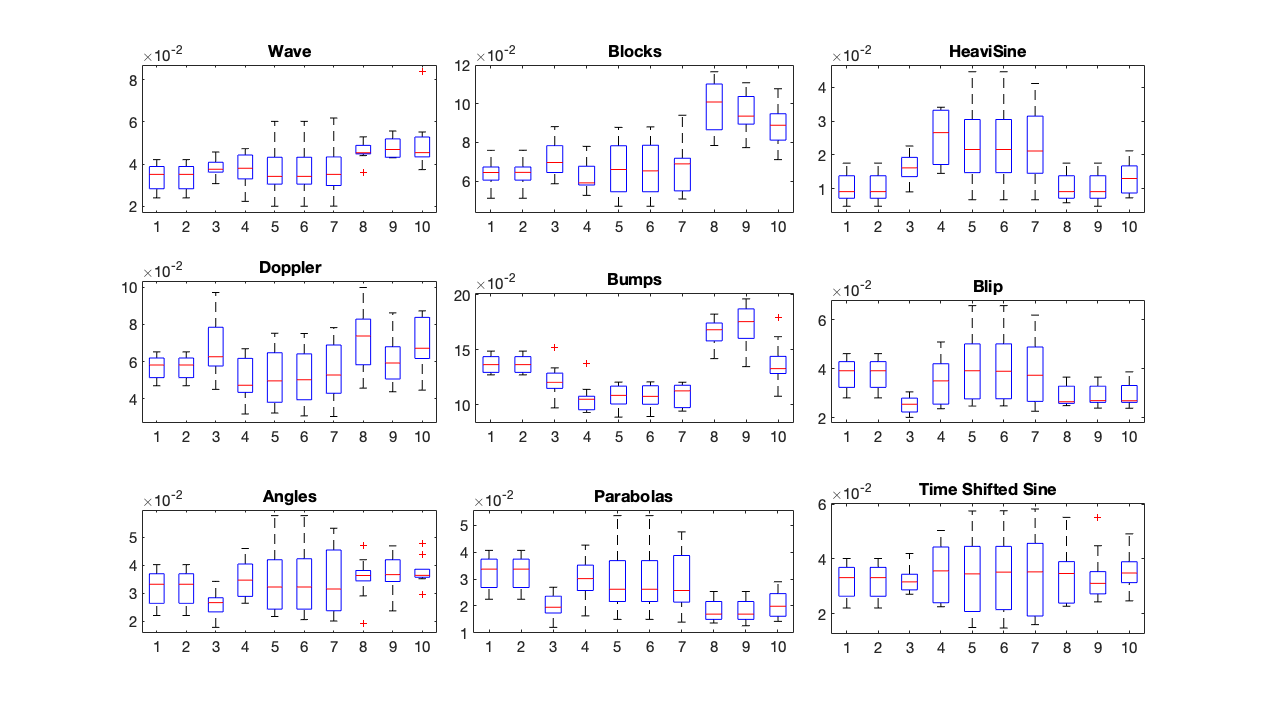}
        \caption{ The box plots of MSE for the ten estimation methods: (1) Rule-I, (2) Rule-II, (3) Bayesian adaptive multiresolution shrinker (BAMS), (4) Decompsh, (5) Block-median, (6) Block-mean, (7) Hybrid version of the block-median procedure, (8) BlockJS, (9) Visu-Shrink, and (10) Generalized cross validation. The MSE was computed by using $SNR= 1/5, k = 2.5, l = 6$, and $n = 1024$ data points.}
\label{fig-5}
\end{figure*}

\begin{figure*}[!t]
\centering
 \includegraphics[width= 1\linewidth]{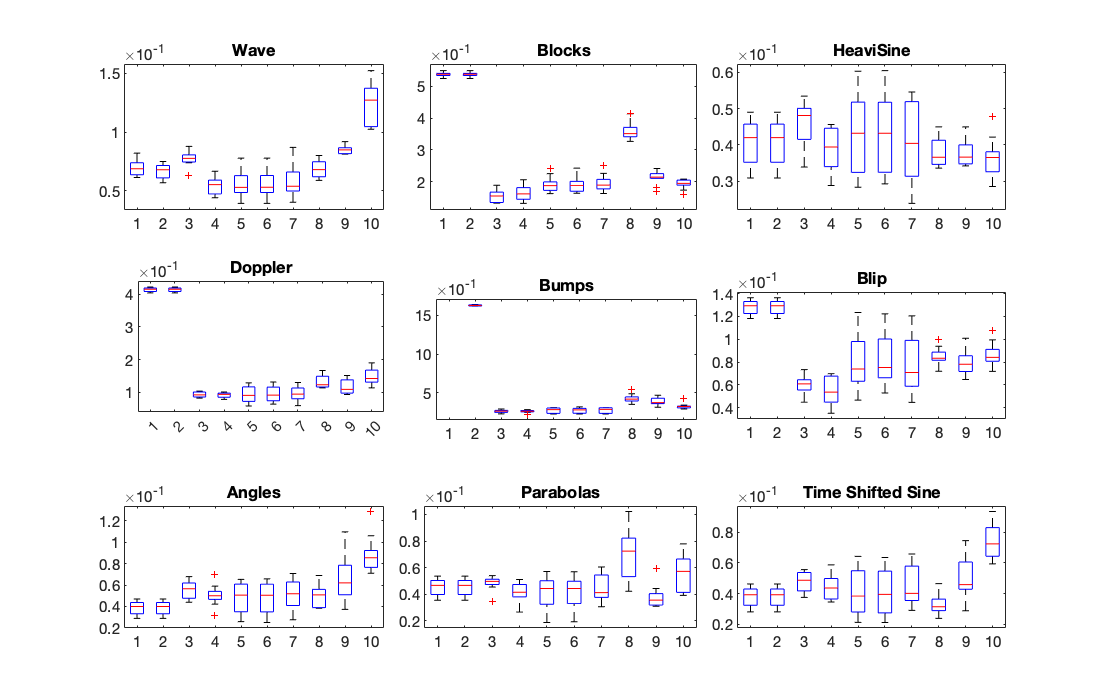}
        \caption{ The same plot as in Fig. \ref{fig-4} with SNR = 3.}
\label{fig-6}
\end{figure*}

In the simulation study, we computed the AMSE using the parameter values of  $l = 6$ and $k =2.5$ and compared with the AMSE computed for the selected estimation techniques.  As can be seen in Fig. \ref{fig-5}, the proposed estimator shows  comparable and for some signals better performance compared to the selected estimation methods. In particular, for smooth signals (e.g., {\tt wave, heavisine}), the three-point prior estimator shows better performance compared to non-smooth signals, such as {\tt blip}, for instance. Moreover, when comparing the performance of the level-dependent estimation methods, the BAMS estimation method shows competitive (or better) performance for most of the cases. We also investigated the influence of SNR level and the sample size on the performance of proposed estimators, compared to the methods considered. For example, for higher SNR (Fig. \ref{fig-6}), the three-point priors-based shrinkage procedure does not provide better performance except for {\tt wave, angels}, and {\tt time shifted sine}. In general, the $\Gamma$-mionimax shrinkage shows comparable or better performance compared to other methods considered, when the SNR is low.  Also, for larger sample  sizes, the three-point prior shrinkage shows improved performance.

\section{Appendix}

\subsection{Proof of Theorem 1}

Consider the class $\Gamma$ of all priors on location $\theta$ in consisting of a point mass at 0 at a fixed level $\epsilon$, and an arbitrary component $\xi(\theta)$ supported on $[-m, m]$ and nonatomic at 0,
\ba
\Gamma =\{\pi(\theta) = \epsilon \delta_0 + (1-\epsilon) \xi(\theta), -m \leq \theta \leq m\}.
\ea
When $\epsilon=0,$ we recover the result from Casella and Strawderman (1981) for which a precise value of $m^*$ is $1.05674351366013496,$ to which we will refer to as Casella-Strawderman's constant.

It is well known result (Levit, 1980; Bickel 1981) that extremal priors in the class of all distributions supported on $[-m, m]$ are symmetric distributions with point masses at 0 and pairs  $-m \leq \pm m_i \leq m, ~i=1,2, \dots$.  Thus, in the $\Gamma$-minimax setup involving this class, the least favorable distributions are Bayes with respect to extremal priors in $\Gamma$, symmetric distributions consisting of point masses.     When $m$ is small (smaller that Casella-Strawderman's constant) the least favorable prior that puts equal weights at the endpoints $\pm m$, that is, the prior $\pi(\theta) = \frac{1}{2} \delta_{-m} + \frac{1}{2} \delta_{-m}$ is the least favorable. The statistical game $\inf_\delta \sup_{\pi} r(\delta, \pi)$ has a value $r(\delta^*, \pi),$ where $\pi$ is the least favorable prior and $\Gamma$-minimax rule $\delta^*$ is Bayes' rule with respect to $\pi.$

In the setup of Model I, the point mass at 0 is a part of every prior; the second component $\xi(\theta)$ is as in Casella and Strawderman (1981), but non-atomic at 0. Conditions of Sion-type theorem, allowing for change of order of inf and sup in (\ref{eq:gmmx}), are not affected by narrowing the class $\Gamma$, the game has value, and for $m < m^*$  the least favorable prior is a three-point mass prior, with masses concentrated at  $-m, 0,$ and $m$, with weights $(1-\epsilon)/2, \epsilon, $ and $(1-\epsilon)/2.$

The corresponding Bayes rule $\delta^*$ is readily found by simplifying
\ba
\delta^*(d) = \frac{\int \theta f(d|\theta)\pi(\theta) d\theta }{\int f(d|\theta)\pi(\theta) d\theta }=
\frac{ (1-\epsilon) m /2 \left[\phi(d|m) - \phi(d|-m)\right]}{ \epsilon \phi(d|0) +  (1-\epsilon)/2 \left[\phi(d|m) + \phi(d|-m)\right]},
\ea
where $\phi(d|\mu)$ is the pdf of normal ${\cal N}(\mu,1)$ distribution. A simplified expression is given in (\ref{eq:breI}).

To find values of $m^*$ so that for $m \leq m^*$ the three point prior is least favorable, we analyze the freqentist risk,
$R(\theta, \delta^*) = E^{d\theta} (\theta - \delta^*)^2,$ for a fixed $\epsilon,$ by varying $m^*.$
Depending on $\epsilon,$ there are three possible shapes of the frequentist risk, which we denote as W, VVV, and V. Numerical work shows that
values of $\epsilon$ that separate these three shapes are $\epsilon_1 \approx 0.45$ and $\epsilon_2 \approx 0.65.$

\begin{figure}[h]
\centering
\threee{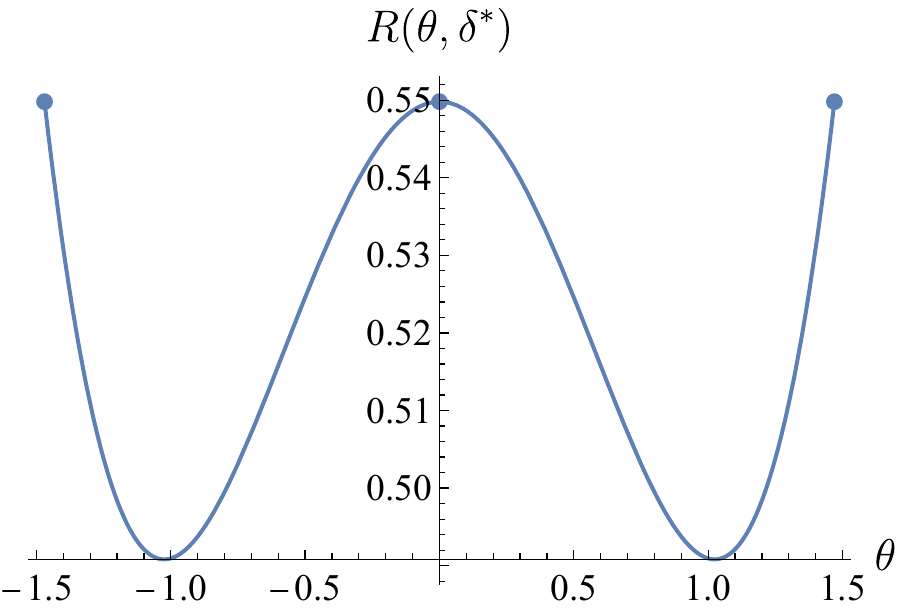}{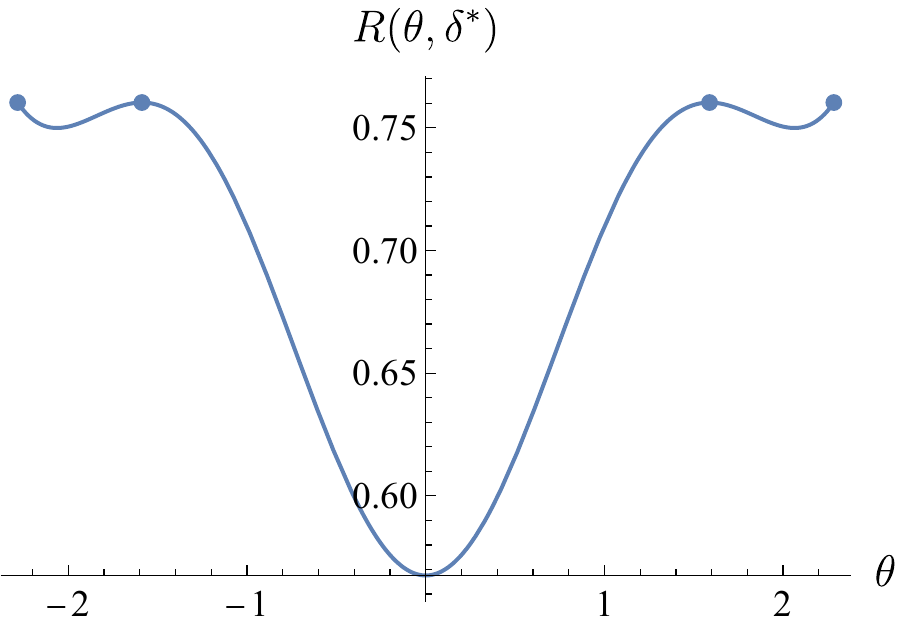}{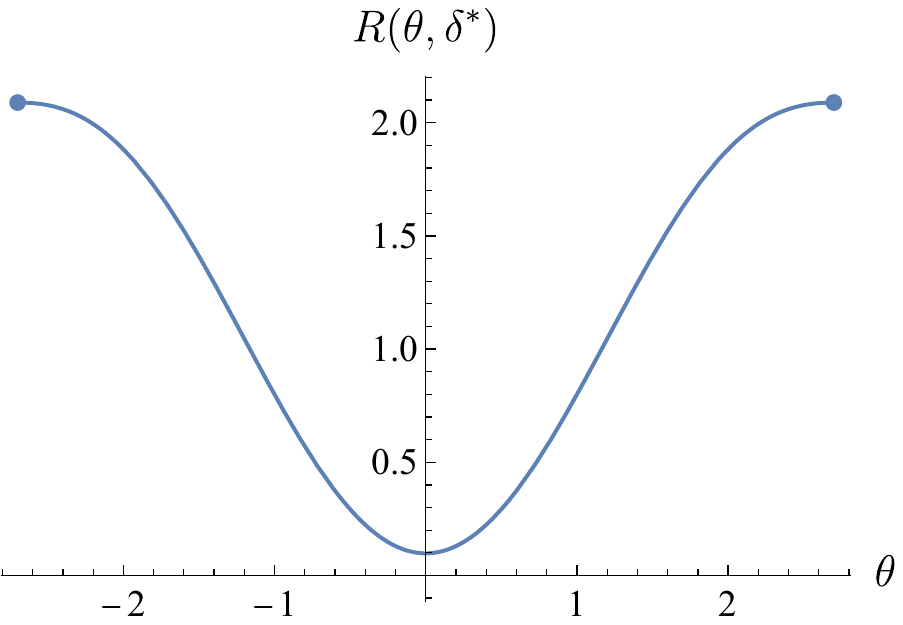}{1.8in}{1.35in}
\caption{Model I: Frequentist risks of W-, VVV-, and V-shape, $R(\theta, \delta^*)$, for $\epsilon=0.3$ (left),
$0.5$ (middle), and $0.9$ (right). \label{fig:riskWI}}
\end{figure}

For small values of $\epsilon$, ($< \epsilon_1$) the risk $R(\theta, \delta^*)$ is of W-shape, as in the left panel of Fig. \ref{fig:riskWI}.

This is a typical shape for a risk of the least favorable distribution in a class of all bounded on $[-m, m]$ distributions, for $m$ small. The value $m^*$ in this case is found form $R(0, \delta*)= R(-m^*, \delta^*) = R(-m^*, \delta^*).$ If $m > m^*,$ the risk local maximum at 0 will exceed values at $\pm m$, and one could select
a prior from $\Gamma$ for which the payoff $r(\delta^*, \pi) > r(\delta^*, \pi_{m^*}.$ Note that in increasing $m$ in the search of this limiting $m^*$, the rule $\delta^*$ is simultaneously changing, since it depends on $m$, so the numerical work to find $m^*$ is nontrivial.

For values of $\epsilon$ between $\epsilon_1$ and $\epsilon_2$, the shape of frequentist risk is of VVV-type, as it is shown in middle panel of Fig. \ref{fig:riskWI}. In this case two local maximums for the frequentist risk appear at
a pair of $\theta=\pm m_1, m_1 \leq m^*$. In the critical case that defines the $m^*$,
$R(-m^*, \delta^*) = R(-m_1, \delta^*) = R(m_1 , \delta^*) = R(m^*, \delta^*)$, and increasing $m$ above such $m^*$ will result in
$R(\pm m^*, \delta_B) < R(\pm m_1, \delta_B).$ Placing more mass at $\pm m_1$ will result in higher payoff $r$ and the three point prior is not the least favorable any longer.

%
%
%

The case when $\epsilon > \epsilon_2$ is most interesting since in the wavelet shrinkage, values of $\epsilon$ closer to 1 produce shrinkage rules of desirable shape. In this case,
 the frequentist risk is V-shaped, which flattens at the endpoints for $m=m^*,$ that is,
~$ \partial R/\partial \theta \left|_{\theta=m^*} = 0\right.$~ (the right panel in Fig. \ref{fig:riskWI}).  In this case if we let $m=m^{**} >m*$ the frequentist risk will start to decrease, so a prior with point masses that remain in now inner points $m^*$ will increase the payoff function $r$, and the three point prior with masses at $0, \pm m^{**}$ will not be the least favorable any longer.

%
%

\subsection{Proof of Theorem 2.}
In Model II the normal likelihood is replaced by double exponential marginal likelihood, after $\sigma^2$ is integrated out.
Meleman and Ritov (1987) show that in estimating bounded normal mean normality of the likelihood is not necessary condition for $\Gamma$-minimax results to remain valid, if mild moment conditions on the likelihood are imposed. In fact they show that if the likelihood has a finite fourth moment, the rescaled asymptotic (when $m\rightarrow\infty$) $\Gamma$-minimax solution has the same least favorable limiting distribution, as for the normal likelihood (Bickel, 1981).

In Model II there are two common shapes of the freqientist risk, W- and VVV-shape, for $\epsilon \leq \epsilon_1$ and $\epsilon > \epsilon_1,$ respectively, with $\epsilon_1$ between 0.3 and 0.4.

\begin{figure}[h]
\centering
\threee{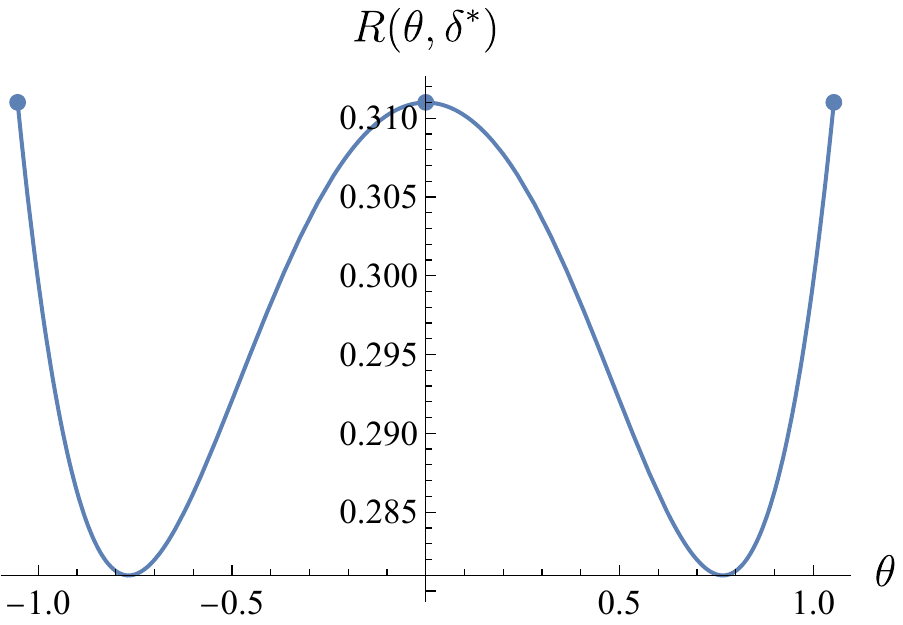}{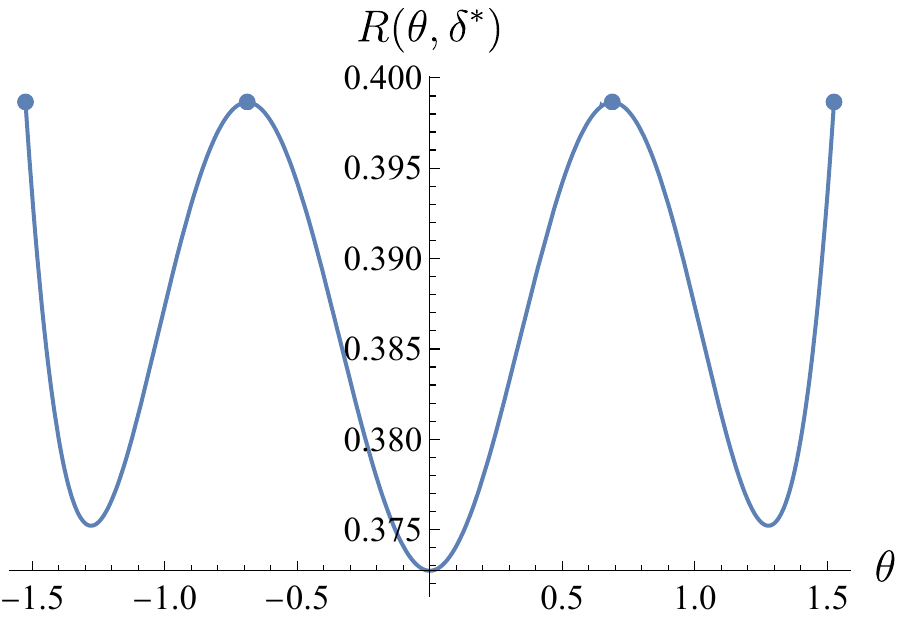}{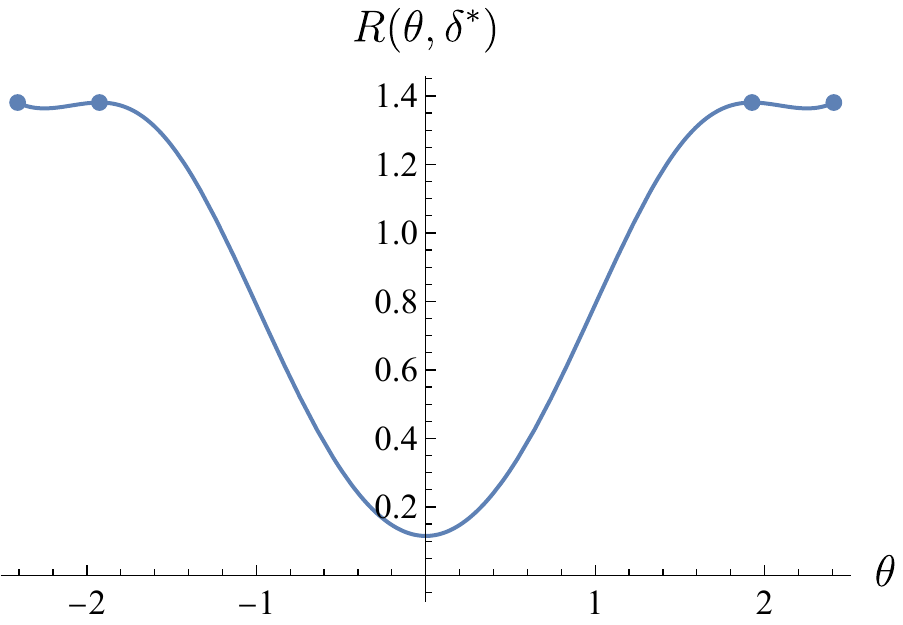}{2.0in}{1.6in}
\caption{Model II: Frequentist risk of the $\Gamma$-minimax rule, $R(\theta, \delta^*)$, for $\epsilon=0.2$ (Left),
$0.4$ (Middle), and $0.9$ (Right).\label{fig:DE} }
\end{figure}

The argument is similar as in Model I, for the W-shape (Left panel in Fig.\ref{fig:DE}), a slight increase of $m$ over $m^*$ will make $R(\delta^*, 0) > R(\delta^*, \pm m)$ and one can choose
$\pm m_1$ in the neighborhood of 0, so that transferring some point mass from endpoints to $\pm m_1$ would increase the payoff. The argument for VVV-shaped risk (Middle and Right Panels in Fig. \ref{fig:DE}) is the same as in the Model I.

%
%
%

%
%

\begin{center}
{\large  REFERENCES}
\end{center}

\leftskip 0.3truein
\parindent -0.3truein


Abramovich, F, Besbeas, P., and Sapatinas, T. (2002). Empirical
Bayes approach to block wavelet function estimation.  {\it Comput.
Statist. Data Anal.}, {\bf 39}, 435--451.

Amato, U. and Vuza, D.T. (1997). Wavelet approximation of a function from samples affected by noise.
{\it Rev. Roumanie Math. Pure Appl.}, {\bf 42}, 481--493.

Angelini, C. and Vidakovic, B. (2004). $\Gamma$-minimax wavelet shrinkage: A robust incorporation
of information about energy of a signal in denoising applications. {\it Stat. Sin.}, {\bf 14}, 1,
103--125.




Antoniadis, A., Bigot, J. and Sapatinas, T. (2001). Wavelet
estimators in nonparametric regression: A comparative simulation
study. {\it  J. Statist. Soft.}, {\bf 6},  1--83.

Berger, J.O. (1984). The robust Bayesian viewpoint. In {\em
Robustness of Bayesian Analysis}, (J. Kadane Eds.) Elsevier
Science Publisher, 63--124.

Berger, J.O. (1985).  {\it Statistical Decision Theory and Bayesian
Analysis}. Springer Verlag, New York.


Bickel, P.J., (1981). Minimax estimation of the mean of a  normal
distribution when the parameter space is restricted. {\it Ann. Statist.},
{\bf 9}, 1301--1309.



Cai, T.T. (1999). Adaptive wavelet estimation: A block thresholding and oracle inequality
approach. {\it Ann.  Statist.}, {\bf 27}, 3, 898--924.

Casella, G. and Strawderman, W. (1981). Estimating a  bounded
normal mean. {\it Ann. Statist.}, {\bf 9}, 870--878.






Donoho, D.L., Liu, R., and MacGibbon, B. (1990). Minimax  risk
over hyperrectangles. {\it Ann. Statist.}, {\bf 18}, 1416-1437.

Donoho, D.L.  and Johnstone, I.M. (1994). Ideal spatial adaptation via wavelet shrinkage. {\it Biometrika}, {\bf 81}, 425--455.




Good, I.J. (1952). Rational decisions. {\it J. Royal Statist. Soc. B},
{\bf 14}, 107--114.

Huang H.C. and Cressie, N., (2000). Deterministic/stochastic wavelet decomposition for recovery of
signal from noisy data. {\it Technometrics}, {\bf 42}, 262--276.



Levit, B. (1980). On asymptotic minimax estimators of second order. {\it Theory Probab. Appl.}, {\bf 25}, 552--568.



Meleman, A. and Ritov, Y. (1987). Minimax estimation of the mean of
a general distribution when the parameter space is restricted.
{\it Ann. Statist.}, {\bf 15},1, 432--442.




Rem\'enyi, N. and Vidakovic, B. (2013).
Bayesian wavelet shrinkage strategies: A review.
In {\it Multiscale Signal Analysis and Modeling,}
Springer,  NY,  317--346.

Robbins, H. (1951). Asymptotically sub-minimax solutions to
compound statistical decision problems. In {\it Proc. Second Berkeley
Symposium Math. Statist. and Prob.}, {\bf 1}, 241-259. 
University of California Berkeley Press.

Sousa, A.R.S., Garcia, N.L., and Vidakovic, B., (2021). Bayesian wavelet
shrinkage with beta priors. {\it Comput. Stat.}, {\bf 36}, 1341--1363.



Vidakovic, B. (2000). $\Gamma$-minimax: A paradigm for
conservative robust Bayesians. In {\em Robust Bayesian Analysis}
(D. Rios Insua and F. Ruggeri, eds.), Lec. Notes Statist. {\bf
152} Springer-Verlag, New York.

Vidakovic, B., and DasGupta, A., (1996). Efficiency of linear rule
for estimating a bounded normal mean. {\it Sankhya  A},  {\bf 58}, 81--100.

Vidakovic, B. and Ruggeri, F. (2001). BAMS Method: Theory and
Simulations.  {\it Sankhya B}, {\bf 63}, 234--249.




 \end{document}